\documentstyle[times]{l-aa}          
\input psfig                    

\def \deg   {$^{\circ}$}

\def \gray     {$\gamma$-ray }
\def \grays    {$\gamma$-rays }

\def \flux     {10$^{-5}$cm$^{-2}$s$^{-1}$}

\begin{document}
 
\thesaurus{   
              03   
              (13.07.02;               
               11.01.2;                
               11.17.4 PKS 0528+134);  
              }

\title{COMPTEL observations of the quasar PKS~0528+134 during the first 3.5 
years of the CGRO mission}


\author{ W.~Collmar\inst{1},
         K.~Bennett\inst{4},
         H.~Bloemen\inst{2},
         J.J.~Blom\inst{2},
         W.~Hermsen\inst{2},
         G.G.~Lichti\inst{1},
         M.~Pohl\inst{1},
         J.~Ryan\inst{3}, 
         V.~Sch\"onfelder\inst{1},
         J.G.~Stacy\inst{3},
         H.~Steinle\inst{1}, 
         and O.R.~Williams\inst{4}
}
 
\institute{  Max-Planck-Institut f\"ur extraterrestrische Physik,
               P.O. Box 1603, 85740 Garching, F.R.G. 
            \and
             SRON-Utrecht, Sorbonnelaan 2, 3584 CA Utrecht,
               The Netherlands
            \and
               University of New Hampshire, Institute for the Study
               of Earth, Oceans and Space, Durham NH 03824, USA
            \and
              Astrophysics Division, Space Science Department of
               ESA/ESTEC, NL-2200 AG Noordwijk, The Netherlands
          }

\offprints{W.~Collmar; wec@mpe-garching.mpg.de}

\date{Received February 15, 1997; accepted July 4, 1997}

\maketitle
\markboth
 {W.~Collmar et al.: COMPTEL observations of the quasar PKS~0528+134 ...}
 {W.~Collmar et al.: COMPTEL observations of the quasar PKS~0528+134 ...}

\begin{abstract}
The COMPTEL observations of the blazar-type quasar PKS~0528+134 in the 
energy range 0.75~MeV to 30~MeV carried out between April 1991 and September 1994 have been analyzed. During the first two years (CGRO Phases I and II), 
PKS~0528+134 was most significantly detected at energies above 3~MeV. During the last year (CGRO Phase III), there is only evidence for the quasar at
energies below 3~MeV indicating a spectral change. 
Detections and non-detections during individual observations indicate a time variable MeV-flux. 
The time-averaged COMPTEL energy spectrum between 0.75~MeV and 30~MeV is well represented by a power-law shape with a photon index $\alpha$ of 1.9$\pm$0.4. However, spectra collected 
from different observational periods reveal different spectral shapes at
MeV-energies: a hard state with $\alpha$=1.4$\pm$0.4 during flaring observations reported by EGRET, and a soft state of $\alpha$=2.6$\pm$0.7 during other
times.    
The combined simultaneous EGRET and COMPTEL spectra indicate these two spectral states as well. During the low luminosity \gray state no spectral break is obvious from the combined COMPTEL and EGRET measurements. Only by inclusion of OSSE data is a spectral bending indicated.
For the high luminosity \gray state however, the combined COMPTEL and EGRET data themselves require a spectral bending at MeV-energies. By fitting a broken power-law shape,
the best-fit values for the break in photon index $\Delta\alpha$ range between 0.6 and 1.7, 
and for the break energy E$_b$ between $\sim$5~MeV and $\sim$20~MeV.
Because the flux values measured by COMPTEL below 3~MeV in both states are roughly equal, the observations would be consistent with an additional spectral component showing up during \gray flaring phases of PKS~0528+134. Such a component could be introduced by e.g. a high-energy electron-positron population with a low-energy cutoff in their bulk Lorentz factor distribution.    
The multiwavelength spectrum of PKS~0528+134 for the \gray flaring phases shows that the major energy release across the entire electro-magnetic spectrum is measured at MeV-energies.

\keywords{ gamma rays: observations; galaxies: active;
           galaxies: quasars: individual: PKS~0528+134 }

\end{abstract} 


 
\section{Introduction}

PKS~0528+134 is a bright radio source with a flat radio spectrum. The polarization was found to be 0.9\%  at 1.465 GHz and 2.2\% at 4.885 GHz (Perley 1982). At higher radio frequencies, PKS~0528+134 shows considerable time variability. Reich et al. (1993) measured a flare during 1992 with an emission increase by a factor of 3.5 at 90 GHz within half a year. An even stronger flare is reported by Pohl et al. (1995) for 1993. 
Recent VLBI images at 22.2~GHz show a one-sided core jet structure of ~$\sim$5~mas length and superluminal motion (Pohl et al. 1995). 
 
In the optical PKS~0528+134 was found to be a $m_v$=20 object with little or no polarization of $0.3\% \pm 1.0\%$ (Fugmann \& Meisenheimer, 1988). An optical spectrum  - obtained in November 1991 with the ESO 3.6m telescope at La Silla/Chile - revealed a redshift of z=2.07, confirming the suspected quasar nature of the source (Hunter et al. 1993). 

At X-rays PKS~0528+134 was detected with the Einstein observatory (Bregman et al. 1985). Recent ROSAT observations by Zhang et al. (1994) and Mukherjee et al. (1996) detected the source. At hard X-rays the source was detected by the OSSE experiment
mainly at energies below 0.5~MeV (McNaron-Brown et al. 1995).

Searches at TeV- and PeV energies for this blazar have been performed by several groups (e.g. Kerrick et al. 1993, Alexandras et al. 1993, Borione et al. 1993). The quasar was detected neither at a few TeV by Cerenkov telescopes nor by any air shower experiment at energies of $\sim$100 TeV or above. 

PKS~0528+134 was detected as a \gray emitter by the EGRET experiment aboard the Compton Gamma-Ray Observatory (CGRO) (Hunter et al. 1993). During two CGRO-pointings towards the galactic anticentre at the beginning of the CGRO-mission, EGRET detected - in addition to the well known \gray sources Geminga and Crab -  a third source. This source was subsequently identified as PKS~0528+134 on the basis of the positional coincidence and its known radio properties.
PKS~0528+134 was detected during several later EGRET observations as well, indicating a variable intensity and spectral behaviour
(Mukherjee et al. 1996).   

The EGRET detections of several blazar type AGN (e.g. Fichtel et al. 1993) during the early CGRO-mission stimulated a search for these AGN in the COMPTEL data. This search led to the discovery of PKS~0528+134 in the COMPTEL data at soft \grays  (0.75 - 30 MeV) by Collmar et al. (1993a). From a preliminary analysis, Collmar et al. (1993b) reported the detection of a spectral break around 10~MeV, by comparing the COMPTEL results of the CGRO viewing period 1 (VP 1) with the contemporaneous EGRET spectrum (Hunter et al. 1993).

\section{Instrument and data analysis}

\subsection{Instrument}

The imaging Compton telescope COMPTEL consists of two layers of detectors. In its primary mode - the double scatter mode - a $\gamma$  photon 
is first Compton scattered in one of the upper detector modules and then - in the ideal case - completely absorbed by a detector module of the lower layer. In this mode, COMPTEL covers the energy range between $\sim$0.75~MeV and $\sim$30~MeV with an energy resolution between 10\% at low energies and 5\% at the high-energy end. Imaging is possible within the 1~steradian field of view with a location accuracy of better than 1\deg ~for sources as strong as the Crab. Different sources of equal intensity  within its field of view can be resolved if they are separated by more than 3\deg - 5\deg. For a detailed description of the COMPTEL instrument see Sch\"onfelder et al. (1993). 
\medskip 

\subsection{Data analysis}         

The COMPTEL data analysis is performed in a 3-dimensional (3d) dataspace, 
defined by 3 quantities: the direction angles $\chi$ and $\psi$  of the Compton scattered photon as projected onto the sky in an arbitrary coordinate system, and the scatter angle $\phi$, which is calculated from the measured energy deposits in the upper and lower COMPTEL detectors. In such a representation, events from a point source will gather along the mantle of a cone, whose apex is pointing in the direction of the source. The COMPTEL analysis software performs a search in this 3d-dataspace for such cone-like structures by applying different methods. The results are source significance maps, flux and flux error maps. In the present analysis, a dataspace binning of 1\deg $\times$ 1\deg
$\times$ 2\deg ~was applied ($\chi$ and $\psi$ corresponding to the galactic coordinates l and b, respectively). 

Usually more than 95\% of the events in this 3d-dataspace are background events (instrumental and from the sky). So for the derivation of source parameters a background estimate has to be generated carefully. The results given in this paper are derived by applying a background generation method which smooths the event distribution in this 3d-dataspace by a running average filter technique. It thereby eliminates - at least to first order - any source signature but keeps the overall event distribution which is considered to be the distribution of the background events. The basic method is described by Bloemen et al. (1994). For consistency checks other background generation methods have also
been applied. 

For mapping and source analysis the COMPTEL analysis system includes the maximum-entropy method (MEM) and the maximum-likelihood method (MLM).
For their 
application to the COMPTEL data see Strong et al. (1992) and de Boer et al. (1992). In the present analysis, the maximum-likelihood method was used to derive detection significances, fluxes and flux errors of possible sources. Images generated by the maximum-entropy method have been used for consistency checks.

The COMPTEL data analysis for the quasar PKS~0528+134 (l:~191.\deg37, 
b:~-11\deg.01) is complicated by its proximity to the Crab (l:~184\deg.56, b:~-5\deg.79), 
which is only $\sim$8\deg ~away and is the brightest source in the COMPTEL energy band. Fig. 1 shows a maximum-likelihood image of the galactic anticentre in the 10-30~MeV energy band. The image is dominated by the Crab though emission from the quasar is clearly evident. From this figure it is apparent that for a parameter estimate of PKS~0528+134, the Crab has to be taken somehow into account. At the upper COMPTEL energy band (10-30~MeV), the two sources are just resolved but towards lower energies the COMPTEL angular resolution decreases. In addition, the Crab outshines the quasar by a factor of about 5 at 10-30~MeV and even more towards lower photon energies.

For determination of the parameters associated with the quasar, we have generated response models of the Crab and PKS~0528+134 and included them 
in the likelihood fitting process, such that their flux values are allowed to vary while their positions are kept fixed. This approach leads to a simultaneous determination of the Crab and PKS~0528+134 fluxes, while generating iteratively  a background model which takes into account the presence of the two sources.
To check for possible systematic effects on the flux values of 
PKS 0528+134 introduced by the proximity of the strong Crab signal, 
we have performed simulations. This is especially necessary for energies
below 3~MeV, for which the Crab signal broadens because the angular resolution
decreases with energy (Sch\"onfelder et al. 1993).
The results of the simulations show that at 
low energies ($<$3 MeV) systematic errors on the PKS 0528+134 fluxes due to 
the proximity of the Crab are $\sim$10\%, which is considerably
smaller than the derived flux errors (see Table 2). 
On the other hand, simulating the Crab and a 'zero-flux' source at the
location of PKS~0528+134 shows that a significant source flux is never obtained
there. This proves that no artifical source is introduced by 
the strong Crab signal. Overall, we estimate systematic errors on 
absolute source fluxes to be up to 30\% due to e.g. imperfections in the
response functions and the background modelling.

The results given in Sect. 4 have been derived with point spread functions (PSFs) assuming an E$^{-2.0}$ power-law shape for both sources. 
The impact of this particular PSF choice on the given flux estimates of PKS~0528+134 is small.

\section{Observations}

Between April 1991 and October 1994, PKS~0528+134 was 11 times in the COMPTEL field of view (within 35\deg ~of the COMPTEL pointing direction). The relevant observational parameters of these pointings are given in Table 1.

For viewing period 0 (CGRO verification phase), we only used data obtained between April 24 and May 7, 1991, when COMPTEL had reached a stable observation configuration. During VP 2.5, a target of opportunity observation of the Sun, COMPTEL was put into a special observing mode to detect solar neutrons. This mode has a reduced efficiency for photon detections. Due to its short duration (1 day) VP 36.0 was always added to VP~36.5. In fact, for final analyses VP~36 and VP~39 have been added together because no hints of the quasar in the individual viewing periods were found.


\begin{table}[hbtp]
\caption[]{COMPTEL observations of PKS~0528+134 during the first 3.5 years of the CGRO-mission. The viewing period (VP) number in CGRO-notation, the observational periods and durations, as well as the angular separation between quasar
position and the COMPTEL pointing direction are given.
 }

\begin{center}\begin{tabular}{cccc}
\hline
  VP   & Obs. Time            & Dur.   &  Ang. Sep.  \\
  \#   & yy/mm/dd - yy/mm/dd  & [days] &  $[ ^{\circ}]$       \\
\hline
  0    & 91/04/22 - 91/05/07 & 15 & $\sim$8  \\
  1    & 91/05/16 - 91/05/30 & 14 & 6.2      \\
  2.5  & 91/06/08 - 91/06/15 &  7 & 5.1      \\
  36.0 & 92/08/11 - 92/08/12 &  1 & 21.2     \\
  36.5 & 92/08/12 - 92/08/20 &  8 & 22.9      \\
  39   & 92/09/01 - 92/09/17 & 16 & 23.9      \\
 213   & 93/03/23 - 93/03/29 &  6 &  9.0      \\
 221   & 93/05/13 - 93/05/24 & 11 &  6.3      \\
 310   & 93/12/01 - 93/12/13 & 12 & 15.6      \\
 321   & 94/02/08 - 94/02/17 &  9 & 12.9      \\
 337   & 94/08/09 - 94/08/29 & 20 & 13.6      \\
\hline
\end{tabular}\end{center}\end{table}


\section{Results}
\subsection{COMPTEL detections}

As a result of a search in the COMPTEL data for EGRET detected blazars, PKS~0528+134 was positively detected during the early part of the CGRO-mission (Collmar et al. 1993a). Fig. 1 shows a maximum likelihood image of the galactic anticentre in the 10-30~MeV band for the sum of VP~0 and VP~1. Apart from the Crab which dominates the map, only at the position of PKS~0528+134 is a feature visible. The log-likelihood ratio value of -2ln$\lambda$=21.8 at the position of the quasar (after correcting for the contribution of the Crab) converts to a detection significance of 4.7$\sigma$ for a known source position ($\chi^2_1$-statistics).


\def\bbllx{ 2.0cm}
\def\bblly{ 8.5cm}
\def\bburx{19.5cm}
\def\bbury{25.9cm}

\begin{figure} [thbp]
\centering{\psfig{figure=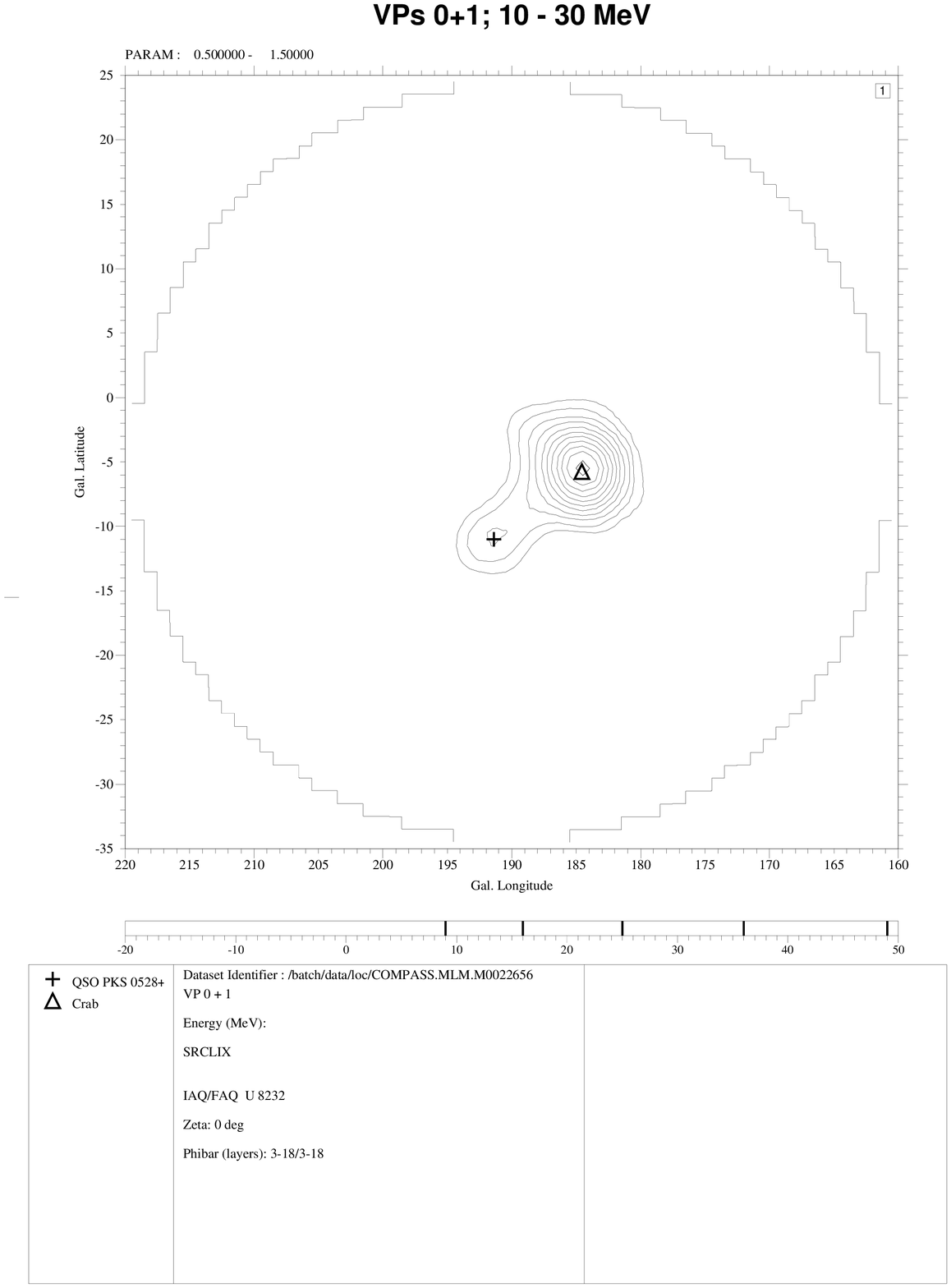,height=7.7cm,bbllx=\bbllx,bblly=\bblly,bburx=\bburx,bbury=\bbury,clip=}}
\caption[]{
COMPTEL 10-30~MeV anticentre map (detection significances) for VP 0 and 1 combined. The contour lines start at a detection significance of 3$\sigma$ (1 d.o.f. for a known source) with a step of 1$\sigma$. The evidence for PKS~0528+134 (+) and the Crab ($\Delta$) is clearly visible.}
\end{figure}


In general, PKS~0528+134 is detected most significantly in the upper 
COMPTEL energy range ($>$ 3~MeV).
Below 3~MeV there are hints in some viewing periods at a significance level up to $\sim$3$\sigma$. However, a note of caution should be added: the COMPTEL point spread function broadens towards lower energies and subsequently the image of the Crab also broadens. Therefore, we expect the systematic uncertainties on the given flux values of PKS~0528+134 below 3~MeV to be larger than above. 
We detect the blazar throughout the entire COMPTEL energy range when we 
take the sum of all viewing periods. Table 2 shows the flux results (detections and upper limits) in the four standard COMPTEL energy bands for all relevant
viewing periods and combinations.

\subsection{Temporal behaviour}

After the early detection in VPs 0 and 1, no hints at all for the quasar could be found during the rest of the CGRO sky survey (VP 2.5, 36, 39). PKS~0528+134 was redetected by COMPTEL during the first observation (VP 213) of the galactic anticentre in Phase II of the CGRO-mission.
The derived fluxes indicate a high intensity level in this viewing period.  A major flare was measured by EGRET at energies above 100~MeV. EGRET observed a factor of 4 flux increase compared to VP~1 (Mukherjee et al.
1996, Fig. 2).  Such an increase in flux is not observed at the highest COMPTEL energies. The 10-30~MeV-fluxes are comparable to the ones measured during VPs 0 and 1. However, a flux increase by a factor 3 to 4 compared to all other observations is observed in the COMPTEL 3-10~MeV band, which indicates that during VP~213 the power-law spectrum measured by EGRET continues further down in energy before it breaks off (Table 2, Fig. 4). During VP~221 - roughly two months later - PKS~0528+134 was not detected anymore.

While up to the end of Phase II of the CGRO mission, PKS~0528+134 was  
detected mainly at the upper COMPTEL energies ($>$3~MeV), during CGRO Phase III there are only indications
(up to $\sim$3$\sigma$) for emission below 3~MeV.  
This behaviour would require a spectral change at MeV-energies. The derived COMPTEL fluxes of CGRO Phase III are consistent
with the OSSE spectrum measured during the same time period (McNaron-Brown et al. 1995).

\begin{table}[hbtp]
\caption[]{
 Upper limits and fluxes for PKS~0528+134 for the different viewing periods. The fluxes are given in units of \flux. The errors are 1$\sigma$. The upper limits are 2$\sigma$. An upper limit is given when the significance of an individual flux value is less than 1$\sigma$. The errors and upper limits are statistical only.}

\begin{center}\begin{tabular}{ccccc}
\hline
  Obs  & \multicolumn{4}{c}{Fluxes [\flux] per Energy band [MeV]} \\
  \#   & 0.75-1 & 1 - 3 & 3 - 10 & 10 - 30  \\
\hline
  0    & 9.4$\pm$8.2  & 9.7$\pm$6.3   & 4.4$\pm$2.5 & 3.2$\pm$1.0 \\
  1    & 16.6$\pm$8.8 & $<$15.7         &  $<$7.2       & 3.0$\pm$1.0 \\
  2.5  & $<$16.2        & $<$21.8         &  $<$9.4       & $<$3.2        \\
36/39  & $<$15.6        & $<$18.7         &  $<$8.4       & $<$2.8        \\
 213   & $<$32.1       & 15.3$\pm$12.3 & 15.2$\pm$4.7 & 3.3$\pm$1.7 \\
 221   & $<$24.4       & 14.0$\pm$9.5  &  4.4$\pm$3.6  & $<$2.7  \\
 310   & $<$20.9       & $<$17.4         &  5.0$\pm$3.6 & $<$4.3        \\
 321  & 11.3$\pm$10.0 & 18.1$\pm$8.3  & $<$6.3        & $<$3.1        \\
 337  & 23.8$\pm$7.9 & 20.6$\pm$6.6  & $<$5.0        & $<$1.8        \\
I+II  & $<$8.7        & 5.8$\pm$3.4    & 4.3$\pm$1.4 & 2.0$\pm$0.5 \\
all   & 4.2$\pm$3.4 & 8.5$\pm$2.7   & 2.3$\pm$1.0 & 1.3$\pm$0.4 \\
high  & 9.1$\pm$5.7 & 6.6$\pm$4.4    & 4.6$\pm$1.7 & 3.1$\pm$0.7 \\
low   & $<$9.8        & 10.5$\pm$3.5  & $<$3.4        & $<$1.1        \\
\hline
\end{tabular}\end{center}\end{table}

The COMPTEL 1-3~MeV and 10-30~MeV light curves together with the EGRET one 
(from Mukherjee et al. 1996) are shown in Fig. 2.
The COMPTEL 10-30~MeV energy band follows the intensity trend as observed by EGRET. PKS~0528+134 is detected by COMPTEL exactly during the periods (CGRO VPs 0, 1, and 213) for which EGRET observes increased source intensity ('flaring'). The quasar is not detected during all other VPs. The COMPTEL 1-3~MeV band however, does not follow the EGRET trend. PKS~0528+134 is best detected during the last two VPs for which EGRET has only marginal detections. 
The detections and non-detections of PKS~0528+134 indicate a time-variable MeV-flux of this quasar. A fit assuming a constant flux of the COMPTEL 
1-3~MeV and 10-30~MeV light curves resulted in $\chi^{2}_{min}$-values of 8.8 and 21.7, respectively. According to $\chi^{2}$-statistics these values correspond to probabilities of 0.36 and 5.6$\times$10$^{-3}$ for a constant MeV-flux or 2.8$\sigma$ evidence for a time-variable 10-30~MeV flux. The fits were done by using the measured flux values which - apart from the detections - scatter around a zero flux level and not by using the upper limits shown in Fig. 2.   
 
\def\bbllx{ 1.0cm}
\def\bblly{12.0cm}
\def\bburx{19.0cm}
\def\bbury{28.0cm}

\begin{figure}[thbp]
\centering{\psfig{figure=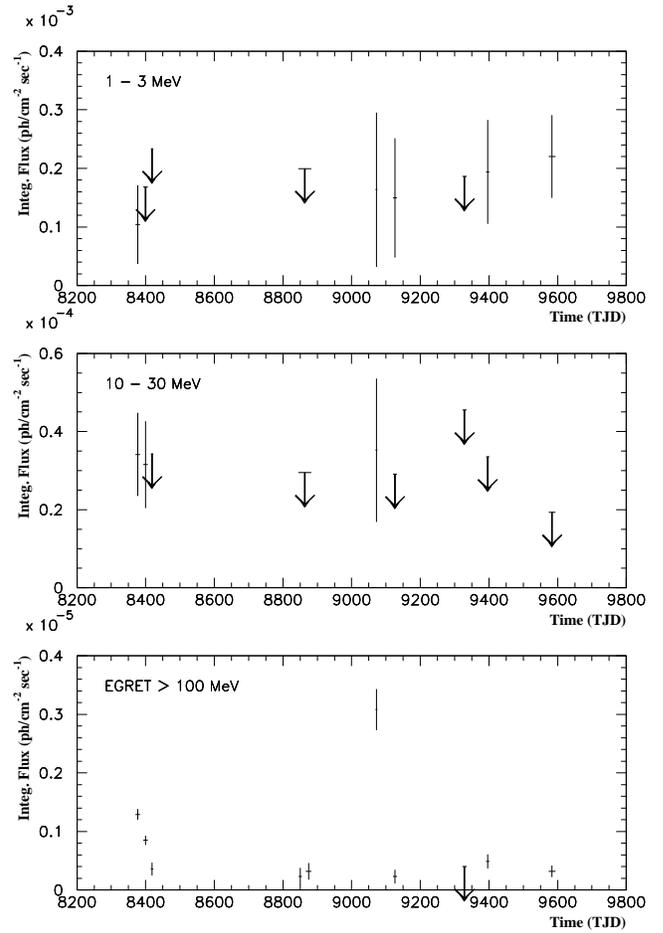,width=8.8cm,bbllx=\bbllx,bblly=\bblly,bburx=\bburx,bbury=\bbury,clip=}}
\vspace{4.3cm}
\caption[]{Time history of PKS~0528+134 as measured by COMPTEL in the 1-3~MeV and 10-30~MeV bands and by EGRET above 100~MeV (from Mukherjee et al. 1996). COMPTEL detects PKS~0528+134 above 10~MeV only for the three EGRET flaring observations. The error bars are 1$\sigma$ and the upper limits are 2$\sigma$. An upper limit is drawn when the significance of an individual flux value is less than 1$\sigma$. The horizontal bars indicate the time intervals covered by the observations.}  
\end{figure}


\subsection{Spectral behaviour}

In order to derive energy spectra of PKS~0528+134 we have applied the maximum-likelihood
method as described in Sect. 2. Background-subtracted, Crab-corrected and deconvolved source fluxes for the four standard COMPTEL energy bands have 
been generated. Table 2 gives the derived flux values for the individual viewing periods and the four standard energy bands. To derive information on the 
spectral shape of PKS~0528+134 in the COMPTEL energy band, we have fitted simple
power-law spectra of the form 

\begin{equation}
I(E) = I_{0}  (E/E_{0})^{-\alpha} \,\, {\rm photons\,\,cm}^{-2} {\rm s}^{-1} {\rm MeV} ^{-1}
\end{equation}  

with the parameters $\alpha$ (photon spectral index) and $I_{0}$ (intensity at the normalization energy $E_{0}$). $E_{0}$ was chosen to be 3~MeV, so that the two free parameters are minimally correlated. 
We derived 1$\sigma$-errors on the parameters by adding 2.3 to the minimum $\chi^{2}$-value (Lampton et al. 1976). The results of the spectral fitting are 
given in Table 3. 

\begin{table}[htbp]
\caption[]{
Results of the powerlaw fitting of the different COMPTEL spectra.
The error bars are 1$\sigma$ ($\chi^{2}_{min}$ + 2.3 for 2 parameters of interest).
 }
\begin{center}\begin{tabular}{cccc}
\hline
  Obs & I$_{0}$ (3~MeV) &  $\alpha$ & $\chi^{2}_{red}$ \\
  \#  & (10$^{-6}$ ph cm$^{-2}$ s$^{-1}$ MeV$^{-1}$   \\
\hline
 all     & 13.8$\pm^{4.6}_{5.1}$ & 1.85$\pm^{0.39}_{0.40}$ & 0.55 \\
 high    & 17.3$\pm^{8.7}_{9.2}$ & 1.38$\pm^{0.34}_{0.49}$ & 0.74  \\
 low     &  8.8$\pm^{5.9}_{6.1}$ & 2.59$\pm^{1.03}_{0.64}$ & 1.04 \\
\hline
\end{tabular}\end{center}\end{table}

Combining all data (labeled 'all' in Tables 2 and 3) we find
an average spectral photon index of -1.85$\pm$0.40 between 0.75~MeV and 30~MeV. Because the temporal analysis indicated different spectral states of the quasar, we have subdivided our data into two parts. Firstly, we have combined the VPs 0, 1, and 213 (labeled 'high' in Tables 2, 3, 4 and in the rest of the paper) for which EGRET has reported high intensity ('flaring') states. Secondly, to increase statistics we have combined all other (low luminosity) observations (labeled 'low' in Tables 2, 3, 4 and in the rest of the paper). The power-law fits to these two spectra resulted in different spectral shapes (Table 3, Fig. 3), indicating different spectral states of PKS~0528+134 at MeV energies. During the high luminosity observations we derive a hard spectrum while during the rest we derive a soft spectrum. The probability that the two spectra are different is 90.5\%. The total, 'low', and 'high' spectra together with their best-fit power-law shapes are shown in Fig. 3. 


\def\bbllx{ 1.0cm}
\def\bblly{12.0cm}
\def\bburx{19.0cm}
\def\bbury{28.0cm}

\begin{figure}[thbp]
\centering{\psfig{figure=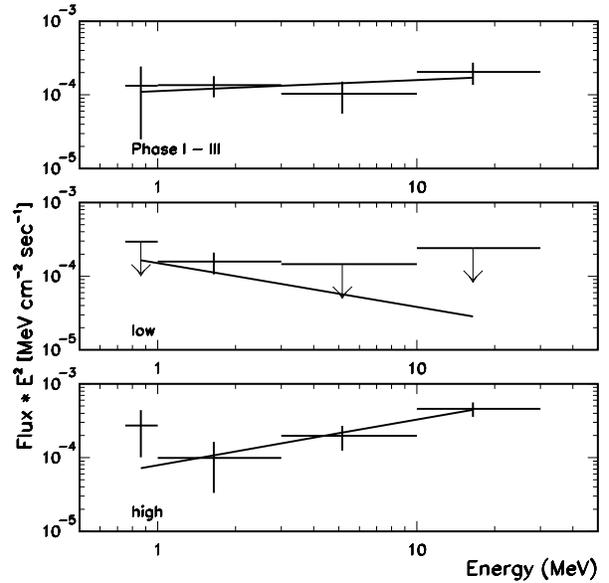,width=8.8cm,bbllx=\bbllx,bblly=\bblly,bburx=\bburx,bbury=\bbury,clip=}}
\caption[]{PKS~0528+134 spectra for all Phase I-III data, the sum of the high  and the low luminosity VPs. The spectra are shown as differential flux $\times$ E$^{2}$. The solid lines represent the best fitting power-law functions. The errorbars are 1$\sigma$ and the upper limits (down arrow) are 2$\sigma$. The fits are performed by using the
measured fluxes and not the upper limits. An upper limit is drawn when the significance of an individual flux value does not reach 1$\sigma$.}
\end{figure}

For weak sources COMPTEL can only derive fluxes in a few, rather broad spectral bands. To constrain any physics, these spectral data have to be compared with results from neighboring energy bands. For blazars we know
that they usually show spectral breaks around MeV-energies e.g. 3C~273 (Williams et al. 1995, Lichti et al. 1995) and PKS 0208-512 (Blom et al. 1995).
Such a behavior had been reported also for PKS~0528+134 (Collmar et al. 1993b, McNaron-Brown et al. 1995). 
To further investigate the spectral behaviour of PKS~0528+134 for 
MeV-energies, we compare the contemporaneously measured EGRET
spectra to the COMPTEL spectra for the different spectral states and for the individual observational periods with significant COMPTEL detections.  
To derive information on the spectral shape 
between 0.75~MeV and $\sim$1~GeV we have performed fits of the
combined COMPTEL and EGRET spectra. The errors on the fit parameters have again
 been
derived by using the method of Lampton et al. (1976). 
Fig. 4 shows the contemporaneously measured
EGRET and COMPTEL spectra of the 'high' and 'low' luminosity combinations as well as of four individual VPs.
The spectra show a bivalent behaviour. While the 'low' state spectra are well represented by simple power-law shapes as usually measured by EGRET at hard \gray energies for blazars (e.g. v. Montigny et al. 1995), the 'high' state spectra require a spectral turnover at MeV energies. In all cases, the extrapolation of the EGRET spectral shape overshootes
the measured COMPTEL fluxes at the lowest energies, indicating a flattening
of the spectrum towards lower energies. Only for VP~213 a simple power-law function results in an acceptable fit. However, the fit improves if a spectral turnover model is applied.  
To derive information on the shape of the flattening we have fitted a broken power-law function of the form 

\begin{equation}
I(E) = \left\{ \begin{array}{ll}
 I_{0}  (E/E_{0})^{-\alpha_{2}} &\mbox {if $E>E_{b}$} \\
 I_{0}  (E_{b}/E_{0})^{-\alpha_{2}} \, (E/E_{b})^{(\Delta\alpha - \alpha_{2})}
                                 &\mbox {if $E<E_{b}$}
\end{array}
\right.
\end{equation}  
where I$_0$ describes the source flux at 200~MeV in units of ph~cm$^{-2}$ s$^{-1}$ MeV$^{-1}$, $\alpha_2$ the high-energy spectral photon index, 
$\Delta\alpha$ the break in spectral photon index towards lower energies
($\Delta\alpha$ = $\alpha_2$ - $\alpha_1$), and E$_{b}$ the break energy. 
To specifically concentrate on the shape and energy of the spectral break
 which are
mainly influenced by the COMPTEL data, the high-energy
power-law index ($\alpha_2$) and the intensity (I$_0$ at 200~MeV) were fixed
in the fitting procedure to the well defined values as derived by fitting  the EGRET spectra only. So, only the two parameters of interest ($\Delta\alpha$, E$_b$) are left as free parameters. Their errors
were derived from the $\chi^{2}_{min}$+2.3 error contours as is appropriate
for two parameters of interest. The fit results are given in Table 4 and the best-fitting broken power-law functions are plotted as solid lines in Fig. 4a-d.

\def\bbllx{ 1.0cm}
\def\bblly{ 2.0cm}
\def\bburx{19.0cm}
\def\bbury{27.0cm}

\begin{figure}[thbp]
\centering{\psfig{figure=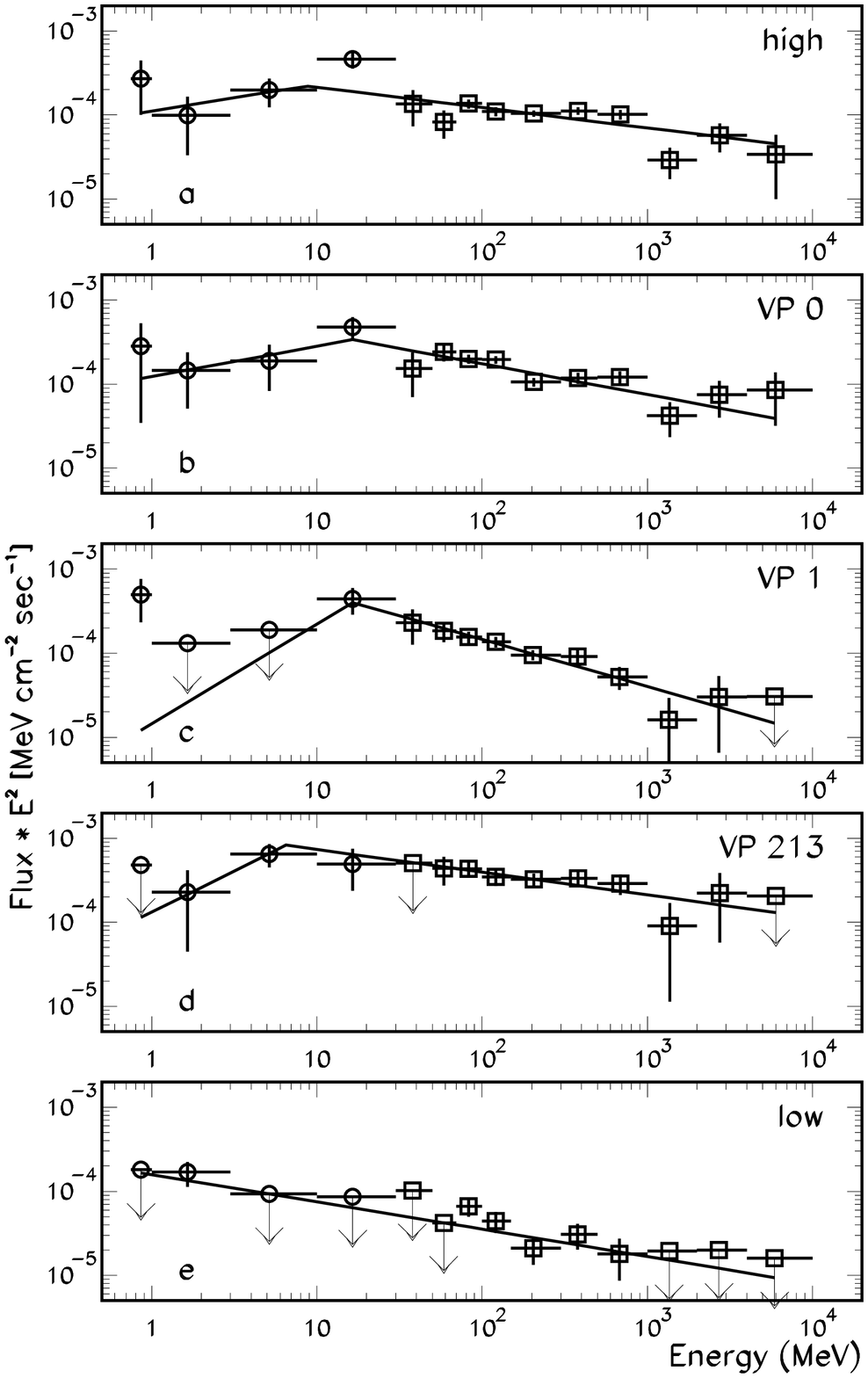,width=8.8cm,bbllx=\bbllx,bblly=\bblly,bburx=\bburx,bbury=\bbury,clip=}}

\def\bbllx{ 1.0cm}
\def\bblly{18.0cm}
\def\bburx{19.0cm}
\def\bbury{29.0cm}
\centering{\psfig{figure=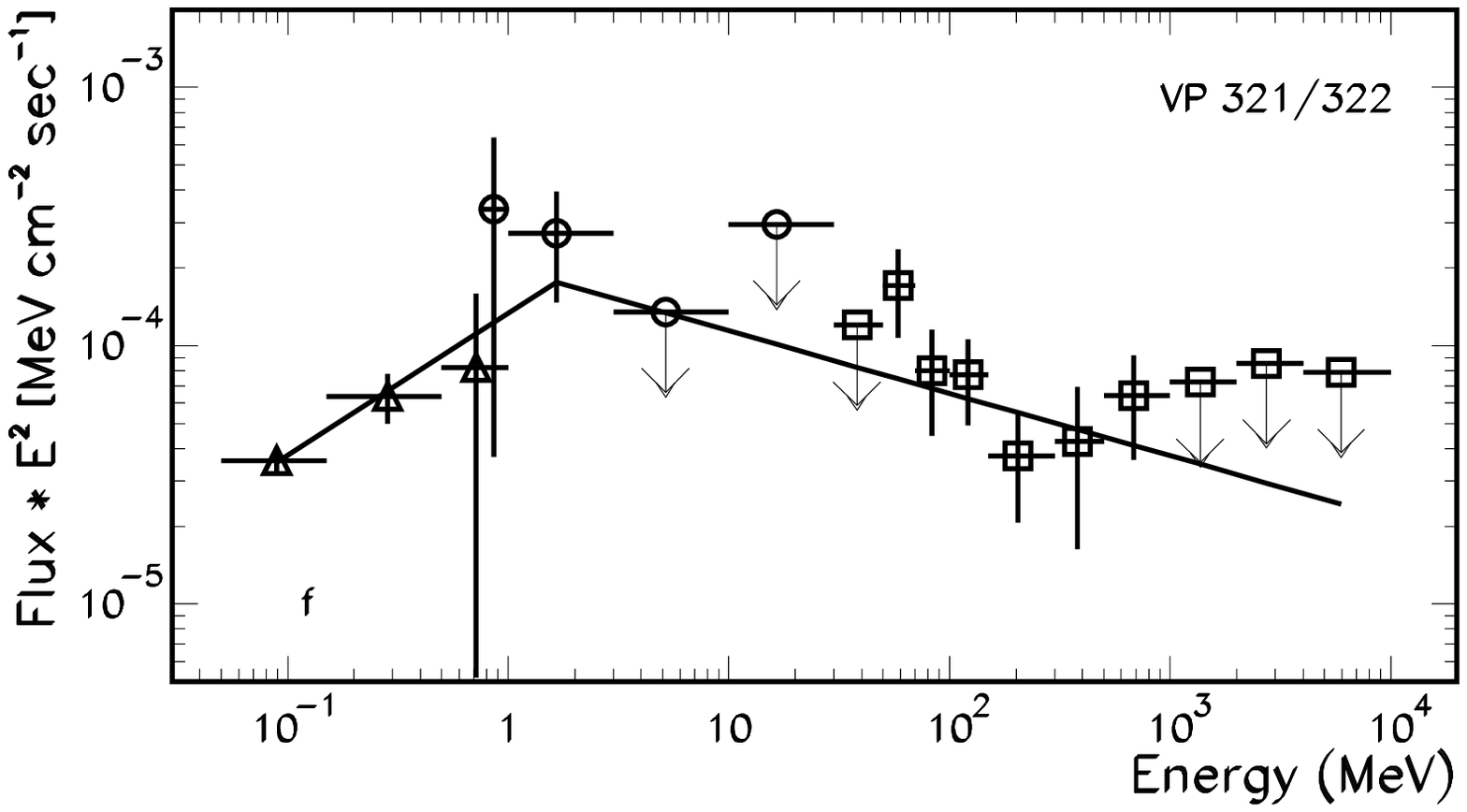,width=8.8cm,bbllx=\bbllx,bblly=\bblly,bburx=\bburx,bbury=\bbury,clip=}}
\caption[]{
Simultaneously measured EGRET and COMPTEL spectra of PKS~0528+134 for the 'high' and 'low' combinations as well as four
individual CGRO viewing periods. The EGRET data are symbolized by $\Box$, the COMPTEL data by $\bigcirc$, and the OSSE data (frame 4f only) by $\triangle$.  The solid lines repesent the best-fitting models. For all 'high' observations (4a-d), the COMPTEL spectral points are not consistent with a simple extrapolation of the measured EGRET spectrum. For both 'low' state observations (4e,f) is a simple power-law shape sufficient from $\sim$1~MeV to $\sim$10~GeV. Only the inclusion of the non-simultaneous OSSE points requires a 
spectral break (4f). Note, that the energy scale in frame 4f is different from the rest. The errorbars as well as the upper limits are 1$\sigma$.
}  
\end{figure}

\begin{table}[htbp]
\caption[]{Results of the spectral fitting for combined EGRET and COMPTEL data measured contemporaneously. The upper part of the table gives the results of the simple power-law fits, while the lower part shows the results of the broken 
power-law fits. In this case I$_{0}$ was fixed at 200~MeV to avoid overlap 
with the break energy in the fitting procedure. The fit improvement by the broken power-law fit is indicated by the reduced $\chi^{2}_{min}$-values.  
 }
\begin{center}\begin{tabular}{cccc}
\hline
 Data & I$_{0}$ (100~MeV)  &  $\alpha$ & $\chi^{2}_{red}$ \\
  \#  & (10$^{-9}$ cm$^{-2}$ s$^{-1}$ MeV$^{-1}$) &  &     \\
\hline
 high   &  11.5$\pm^{0.9}_{0.9}$ & 2.18$\pm^{0.04}_{0.05}$ & 3.03  \\
 low    &   3.6$\pm^{0.8}_{0.8}$ & 2.33$\pm^{0.10}_{0.11}$ & 0.82  \\
VP 0    &  15.0$\pm^{1.3}_{1.3}$ & 2.20$\pm^{0.05}_{0.05}$ & 2.58  \\
VP 1    &  11.0$\pm^{1.3}_{1.3}$ & 2.27$\pm^{0.05}_{0.06}$ & 2.87  \\
VP 213  &  33.2$\pm^{4.6}_{4.6}$ & 2.11$\pm^{0.06}_{0.07}$ & 1.04  \\
VP 321/322 &  5.2$\pm^{1.2}_{1.7}$ & 1.94$\pm^{0.08}_{0.02}$ & 1.13  \\
\hline
\end{tabular}\end{center}

\begin{center}\begin{tabular}{cccccc}
\hline
 Data  & I$_{0}$  & $\alpha_2$ & $\Delta\alpha$ & E$_{b}$ & $\chi^{2}_{red}$ \\
  \#   & fixed    &  fixed     &                &  (MeV)  &                  \\
\hline
high & 2.60 & 2.24 & 0.56$\pm^{1.42}_{0.39}$ &  8.8$\pm^{31.0}_{7.3}$ & 2.43  \\
VP 0 & 3.38 & 2.37 & 0.72$\pm^{1.39}_{0.45}$ & 16.5$\pm^{67.4}_{9.9}$ & 1.47  \\
VP 1 & 2.47 & 2.56 & 1.75$\pm^{\infty}_{0.92}$ & 16.5$\pm^{23.4}_{10.3}$ & 0.67  \\
VP 213& 8.18& 2.27 & 1.24$\pm^{\infty}_{0.94}$ &  6.5$\pm^{63.9}_{6.3}$ & 0.42  \\
\hline
VP 321/2 & 1.39$\pm^{0.45}_{0.45}$ & 2.24$\pm^{0.21}_{0.19}$ & 0.75$\pm^{0.44}_{0.25}$ & 1.7$\pm^{15.8}_{1.2}$ & 0.58  \\
\hline
\end{tabular}\end{center}
\end{table}

With respect to the simple power-laws, the broken power-law functions improve the fits. For all four cases, the best-fit values for the break in photon index are larger than 0.5. They range between $\sim$0.6 and $\sim$1.7. However, only for VP~1 is the value of 0.5 excluded by the 
1$\sigma$ error contour (Fig. 5). A break is always required, because the 1$\sigma$ error contours do not reach a value
of 0 for $\Delta\alpha$. Upper limits on $\Delta\alpha$ are not
 defined for VP~1 and VP~213
because the lower-energy COMPTEL points are not significant enough to constrain the shape for large $\Delta\alpha$- and small E$_b$-values.
Best-fit values for the break energy range between $\sim$5~MeV and $\sim$20~MeV
(with large error bars). A correlation plot of $\Delta\alpha$ versus E$_b$ for VP~1 shows that the two parameters are highly 
correlated and that a break value of 0.5 is excluded by the 2$\sigma$ error contour.

For VP~322 (April 5-19, 1994) OSSE reports a strong detection of PKS~0528+134
with a hard photon spectrum ($\sim$E$^{-1.2}$) between 50~keV and $\sim$1~MeV
(McNaron-Brown et al. 1995). VP~322 is spaced in time between the VPs 321 and
337 for which 
COMPTEL detects the quasar only at energies below 3~MeV (Table 2). In order 
to investigate the spectral shape of PKS~0528+134 at $\gamma$-rays for this
time period, we generated a combined OSSE/COMPTEL/EGRET-spectrum. We used the  
OSSE fluxes of VP~322 given by McNaron-Brown et al. together with the 
COMPTEL- and EGRET-data of VP~321 ('low' state) which took place about two months earlier and is the closest COMPTEL/EGRET observational period.
The combined COMPTEL/EGRET spectrum for VP 321 is a
typical 'low' state spectrum  and consistent with a simple power-law shape. However, the inclusion of the OSSE points, which are highly 
significant ($\sim$10$\sigma$) at energies below 0.5~MeV,
requires a spectral break around the lowest COMPTEL energies. 
The combined OSSE/COMPTEL/EGRET-spectrum together with the best-fitting 
broken power-law function is shown in Fig. 4f. The best-fit values for 
single and broken power-law fits are given in Table 4. 
In contrast to the combined COMPTEL/EGRET fits discussed above, we did not fix 
any parameter in the broken power-law fit. Due to the enlarged spectral lever-arm towards lower energies 
a fixing of the parameters was not required. 
The simple power-law fits converged with  a statistically sufficient reduced
$\chi^{2}_{min}$-value of 1.13. However, the improvement in $\chi^{2}_{min}$
of 9.4 for a broken power-law fit converts to a probability of 99.1\% 
(2.6$\sigma$) that the broken power-law shape is a better representation of the measured data. Again, the best-fit value of 0.75 for $\Delta\alpha$ is larger
than 0.5, but is consistent with 0.5 at the 1$\sigma$-level. For the 
break energy we derive a best-fit value of 1.7~MeV which is the lowest value
derived for this parameter. Note however, that in contrast to the combined COMPTEL/EGRET results this combined OSSE/COMPTEL/EGRET result is derived
from non-simultaneous data.

\begin{figure} [thbp]
\centering{\psfig{figure=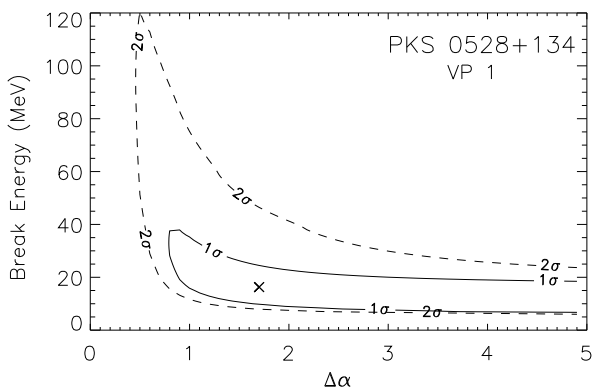,width=8.8cm,clip=}}
\caption[]{
Correlation between the break in photon index ($\Delta\alpha$) and the break energy for VP~1. The best-fit value (x) as well as the $\chi^2_{min}$+2.3 (1$\sigma$) and $\chi^2_{min}$+6.2 (2$\sigma$) contour levels are shown.
The 1$\sigma$-contour clearly excludes a break value of 0.5 while the 
2$\sigma$ just reaches this value. Due to the small significance below 10~MeV
an upper limit on $\Delta\alpha$ can not be derived using equation 2.  
}  
\end{figure}


Fig. 6 shows a broad-band spectrum of PKS~0528+134. The data are selected to be as close as possible to the April 1991 CGRO-observation (VP~0). However, only the EGRET and COMPTEL data are contemporaneous. It is seen that the radiated power per logarithmic frequency interval during this time period peaks at MeV-energies. This is also the case for at least all 'high' state observational periods (Fig. 4).    


\def\bbllx{ 1.0cm}
\def\bblly{14.0cm}
\def\bburx{19.0cm}
\def\bbury{28.0cm}

\begin{figure}[thbp]
\centering{\psfig{figure=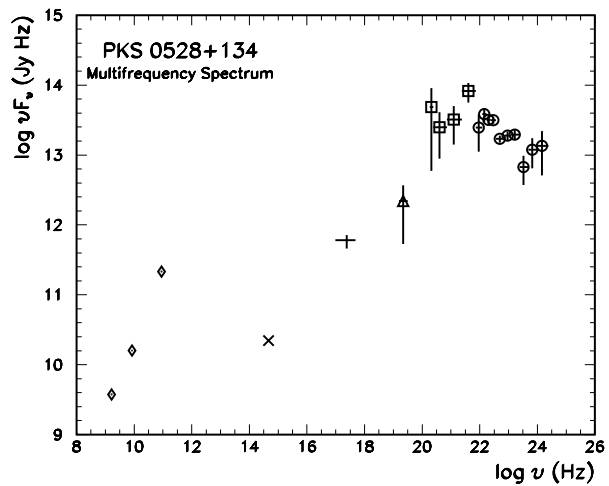,width=8.8cm,bbllx=\bbllx,bblly=\bblly,bburx=\bburx,bbury=\bbury,clip=}}
\caption[]{
Broad band spectrum of PKS~0528+134 for VP~0 (April 1991). Only the COMPTEL ($\Box$) and EGRET-data ($\bigcirc$) are measured contemporaneously. Apart from the OSSE-measurement ($\triangle$) (October 1992, McNaron-Brown et al. 1995) the other results are measured within three month of the CGRO-observation. The radio data ($\Diamond$) are taken from Reich et al. (1993). The soft X-ray (ROSAT) point (+) is from Zhang et al. (1994), and the hard X-ray point (OSSE) is from McNaron-Brown et al. (1995). The optical point (x) was supplied by Wagner (1995, priv. comm.).     
}  
\end{figure}


\section{Discussion}

With the detection of more than 60 active galactic nuclei (AGN) at $\gamma$-ray energies (Kanbach 1996), EGRET opened the field of extragalactic \gray astronomy. Common characteristics of these blazar-type AGN are: (1) strong radio emission from compact regions with a flat radio spectra, (2) variability in the optical and IR, (3) superluminal motion (indicative of jets) for a significant fraction, and (4) time variability on time scales from days to months (or even longer) in \grays.
Since their discovery, the origin of the \gray emission from these sources has been widely discussed. Isotropical central core emission has to be excluded. 
The compactness of the emission region inferred from the observed \gray flux and time variability would lead to severe absorption of the \gray radiation by
photon-photon interactions and therefore would contradict  
the observed power-law emission into the GeV-range. To avoid this contradiction, models have focused on a relativistic jet scenario in which the \gray emission is directed close to the observer's line of sight (beaming), thus reducing the required internal \gray luminosity by several orders of magnitude, due to relativistic Doppler boosting of the emitted photons and the solid angle effect. Most prominent are the inverse Comptonization (IC) models in which
low-energy photons are scattered to $\gamma$-energies by the relativistic 
jet electrons. These models can be subdivided whether self generated synchrotron photons (SSC process; e.g. Maraschi et al. 1992, Bloom \& Marscher 1993) or external photons (external radiation Compton process or ERC process; e.g. Dermer \& Schlickeiser 1993, Sikora et al. 1994, Blandford \& Levinson 1995) dominate in the scattering process.
These models make predictions about the spectral shape of blazars in the \gray range. For example, the ERC models by Dermer \& Schlickeiser (1993) and
Sikora et al. (1994) predict a spectral break in power-law index of $\Delta\alpha \sim$ 0.5 at MeV-energies due to incomplete Compton cooling of 
the electrons. However, Dermer et al. 1997 showed that pure ERC models are
able to produce spectral breaks larger than 0.5 as well by considering 
further cooling processes like Coulomb cooling for example.  

A separate class of models considers the emission from a relativistic pair
plasma in more detail (e.g. Henri et al. 1993,
Marcowith et al. 1995, Blandford \& Levinson 1995, B\"ottcher \& Schlickeiser
1996). In addition to IC scattering of soft photons as in the SSC- and
ERC-model, which is still the dominant process, pair-annihilation and 
pair-cascades are considered. This could result in distinct spectral features
at MeV energies like spectral humps due to blue-shifted annihilation
radiation and large spectral breaks/cutoffs ($>$0.5) due to $\gamma$-$\gamma$
pair attenuation above the pair-production threshold.

\subsection{Black Hole mass and \gray transparency}

During these 3.5 years of its mission,
COMPTEL has observed time-variable MeV-emission from PKS~0528+134 with the largest flux value observed in March 1993 for energies above 3~MeV. Using this measured 3-30~MeV flux and assuming isotropic emission, we derive a luminosity of $\sim$4.1$\times 10^{49}$erg/s for z=2.07 (Hunter et al. 1993), H$_0$=60~km/s/Mpc, and a q$_0$=0.5 cosmology. This is the largest luminosity yet derived for any blazar. The corresponding minimum black hole mass for Eddington-limited accretion in the Thomson regime (M$_8^T>$L$_T$/(1.26$\times$10$^{46}$erg/s) would be $\sim$3.2$\times$10$^{11}$M$_\odot$. Dermer \& Gehrels (1995) point out however,
 that for photon energies larger than 0.511~MeV Klein-Nishina effects on the Compton-scattering cross section have to be considered when inferring Edington-limited masses. To calculate the minimum black hole mass in units of 10$^8$M$_{\odot}$ for this energy range, assuming isotropic radiation and Eddington-limited accretion, Dermer and Gehrels provide the following expression 
\begin{equation}
M_{8} \geq \frac{4\pi d^2_L(m_ec^2)(1+z)^{-2}}{1.26 \times 10^{46}ergs s^{-1}} \int_{\epsilon_l(1+z)}^{\epsilon_u(1+z)} \epsilon \Phi (\frac{\epsilon}{1+z}) \kappa (\epsilon) d\epsilon
\end{equation}

where d$_L$ is the luminosity distance,
z is the redshift of the source, $\epsilon$ is the dimensionless photon energy in units of m$_{e}$c$^2$, $\epsilon_{l,u}$ the lower and upper limits on the observed energy range, $\Phi$ the differential photon flux, and $\kappa(\epsilon)$ is a function describing the Compton radiation force. 
Integrating this equation  
between 3 and 30~MeV using the measured COMPTEL fluxes we derive a minimum black hole mass of $\sim$85$\times$10$^{8}$M$_\odot$, which leads to a minimal Schwarzschild radius (R$_S$ = 2GM/c$^2$) for the putative black hole of $\sim$2.55$\times$10$^{15}$cm or roughly 1 light-day. COMPTEL can claim variability for PKS~0528+134 between this March '93 and the May '93 observational periods (Tables 1,2). The observations are 54 days apart which means $\sim$18 days in the source frame (z=2.07). Because these 18 days are larger than the inferred Schwarzschild radius in units of light-days, no beaming of the MeV-photons is required according to this so-called Elliot-Shapiro relation (Elliot \& Shapiro 1974). 

If one assumes that the X-rays and \grays are produced cospatially the \grays will be absorbed by $\gamma$-ray/X-ray pair production.
 Applying the expressions of 
Mattox et al. (1997) we derive an optical depth $\tau_{\gamma\gamma}$ of 10 and a doppler boost factor of $\geq$1.5. For the X-ray $\nu$F$_{\nu}$-flux at 1~keV  we have taken the results of a ROSAT observation of PKS~0528+134 in September 1992 reported by Mukherjee et al. (1996). A value of $\tau_{\gamma\gamma}$ of $\sim$10 together with the measured COMPTEL and EGRET spectra excludes an isotropic \gray emission and therefore implies a beamed \gray emission. This result, however, still contains the unproven assumption that the X-rays and \grays are generated at the same time at the same source location. We can avoid the need of this assumption if we just consider the \gray photons observed in the 3-30~MeV COMPTEL energy range themselves. Dermer (1997) gives an equation 
for the pair production optical depth $\tau_{\gamma\gamma}$ for a photon with 
observed dimensionless energy $\epsilon^{obs}_1$ in units of m$_{e}$c$^2$

\begin{equation}
 \tau_{\gamma\gamma}(\epsilon^{obs}_1) \simeq \frac{3f^2d^2_L}{c^2 \delta t_{obs}}
 \int_{max[2/[\epsilon^{obs}_1(1+z)^2],\epsilon^{obs}_l]}^{\epsilon^{obs}_u} \sigma_{\gamma\gamma}(\tilde{s}) \Phi(u) du
\end{equation}

where $\epsilon^{obs}_{l,u}$ are the lower and upper limits on the observed
energy range, u=$\epsilon$(1+z), $\sigma_{\gamma\gamma}$ is the 
pair-production cross section, $\tilde{s}\equiv$2(1+z)$^2$u$\epsilon^{obs}_1$, $\Phi(u)$ is the differential flux, d$_L$ is the 
luminosity distance, $\delta t_{obs}$ is the observed variability, and f a factor describing the mean distance from the inside to the boundary of an emitting sphere. 
Integrating this equation with the measured COMPTEL data, we derive a  $\tau_{\gamma\gamma}$(3~MeV) of $\sim$2.4. This value, together with the measured COMPTEL flux below 3~MeV hints for beaming of the MeV-emission
observed by COMPTEL.

\subsection{Spectra and time variability}

COMPTEL has detected the quasar in all 4 standard energy bands at least once near or above the 3$\sigma$-level.
The average COMPTEL Phase I-III spectra (Fig. 3; Table 3) are consistent with power-law representations with a photon spectral index of $\sim$1.9 which is harder than the spectral indices between $\sim$2.2 and $\sim$2.6 as measured by EGRET (Mukherjee et al. 1996) and softer than the index of 1.2 as measured by OSSE (McNaron-Brown et al. 1995). This indicates that the COMPTEL energy range is a transition region for the spectrum of this blazar. 
 Subdividing the COMPTEL observations into 'high' and 'low' viewing periods we derive different spectral states: a hard state for the EGRET flare phases and a soft state otherwise. While below $\sim$3~MeV both flux values are comparable, they are quite different at the 10-30~MeV range.

Like the COMPTEL spectra themselves, also the combined COMPTEL/EGRET spectra show a 'two state' behaviour. Both 'low' state spectra investigated are consistent with a simple power-law fit for MeV- to GeV-energies.
For the 'high' state spectra investigated, however, is no simple extrapolation of the EGRET spectrum throughout the COMPTEL energy range possible (Fig. 4, Table 4). Broken power-law fits represent the data much better than simple power-law representations giving the evidence for a spectral turnover
within the COMPTEL energy range. The best-fit break values are always larger than the value of 0.5 as predicted by different ERC models (Dermer \& Schlickeiser 1993, Sikora et al. 1994). However, only for VP 1 a break value of 0.5 is excluded at about the 2$\sigma$-level. 

If we combine the published OSSE spectral points of VP 322 (McNaron-Brown et al. 1995) with the COMPTEL/EGRET ones of VP~321 ('low state') a spectral turnover at the lower edge of the COMPTEL energy range becomes evident. The derived break value of 0.75 is just at the 1$\sigma$-level consistent with the predicted value of 0.5 by the ERC models. The observed 'low' state' spectra are basically consistent with the simple versions of e.g. the ERC- and SSC-models. The measured spectrum resembles the spectrum of 3C~273, where a steepening of $\sim$0.8 for E$\gg$1~MeV was found (Lichti et al. 1995). 


\def\bbllx{ 1.0cm}
\def\bblly{14.0cm}
\def\bburx{19.0cm}
\def\bbury{28.0cm}

\begin{figure}[tbph]
\centering{\psfig{figure=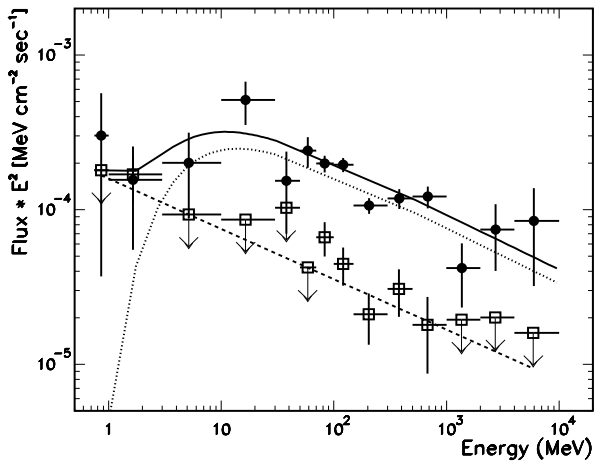,width=8.8cm,bbllx=\bbllx,bblly=\bblly,bburx=\bburx,bbury=\bbury,clip=}}
\caption[]{PKS~0528+134 COMPTEL/EGRET spectra for the low \gray luminosity ('quiescent') observations (open squares) and VP~0 (filled circles). The dashed line represents the best-fit power-law shape for the 'low' state spectrum. The dotted line symbolizes the best-fit EGRET power-law-spectrum for VP~0 with an exponential low-energy cutoff added however. The solid line shows the sum of both. The error bars and upper limits (down arrow) are 1$\sigma$. For the interpretation see the text.}
\end{figure}

The combined spectra generated for 'quiescent' and 'flaring' \gray emission suggest that during flares in the EGRET energy range an additional spectral component is showing up which cuts off from a power-law shape around the lowest EGRET energies ($\sim$50~MeV) and reaches down to the middle of the COMPTEL energy range around 3~MeV. Such an additional spectral component is suggested by three observational facts: (1) the COMPTEL light curve above 10~MeV clearly follows the flux trend of PKS~0528+134 as observed by EGRET at energies above 100~MeV, but does not so for \gray energies below 3~MeV, (2) the COMPTEL 'high' and 'low' state spectra have different spectral slopes where the fluxes measured below 3~MeV are the same within error bars, and (3) the combined COMPTEL/EGRET spectra for the low state observations require no spectral break while the flaring state observations always require a spectral break with best-fit values above $\sim$5~MeV. Fig. 7 illustrates our interpretation of the data on PKS~0528+134. It shows the spectral points and the best-fit power-law fit (dashed line) for the low luminosity ('quiescent') observations. Above we show the combined COMPTEL/EGRET spectrum of VP~0. The dotted line represents the power-law shape measured solely by EGRET (30~MeV - 10~GeV) for this observation, however, with an exponential low-energy cut off added. If we simply add both shapes ('quiescent' and 'flaring') we obtain the solid line, which is in qualitative agreement with our measurement. We note that this solid line is not a fit to the data. We just want to illustrate that an additional spectral component with a low-energy cutoff sitting on top of a quiescent level can generally describe the combined COMPTEL/EGRET spectra for such flaring observations. We also like to note that the assumed exponential shape for this cutoff is arbitrary. A broken power-law function as fitted in Sect.~4 would have illustrated this effect as well. Given the bandwidth and the significance of the spectral points, we can not distinguish between different cutoff shapes. 

Our data are consistent with a two component \gray spectrum of PKS~0528+134
during flaring periods: an additional high-energy component on top of 
a quiescent \gray component. A possible explanation for such a scenario is
provided by the model of B\"ottcher et al. (1997), which considers the ejection of an electron pair population with a low-energy cutoff in its Lorentz-factor distribution into the emission region. According to their model, protons - accelerated to the thresholds for photo-pair and photo-pion production and the threshold for pion production in inelastic proton-matter collisions - will generate plenty of secondary electrons and positrons of ultrahigh energies which are now injected into the acceleration scheme. If these secondary particles are generated outside the \gray photosphere, the secondary electrons and positrons will quickly cool by strong inverse-Compton losses resulting in a cooling particle distribution with a strong cutoff towards lower bulk Lorentz factors. Inverse-Compton scattering of external e.g. accretion disc radiation by such a particle population would result in an additional spectral component showing up between MeV- and GeV-energies in time-integrated \gray blazar
spectra (B\"ottcher et al. 1997).

The high state spectra are also consistent with the predictions of the
pair-plasma models (e.g. Henri et al. 1993,
Marcowith et al. 1995, Blandford \& Levinson 1995, B\"ottcher \& Schlickeiser
1996).  A spectral break is observed
(at least once significantly larger than 0.5) with an energy spectral
index above the break consistent with a value of twice the one below the break, 
as is predicted by Henri et al. (1993). However, no obvious spectral 'bump'
at MeV-energies indicative of an annihilation line feature is observed. 
The upper COMPTEL spectral points agree well with the extrapolation of the 
EGRET spectra (Fig. 4). This is different to the spectral 'humps' observed 
by COMPTEL at MeV-energies from two so-called MeV-blazars
(Bloemen et al. 1995, Blom et al. 1996), which are interpreted as the
signature of a broad blue-shifted annihilation line (Roland \& Hermsen 1995).

\section{Conclusion}

The COMPTEL observations of the quasar PKS~0528+134 between April 1991 and September 1994 (CGRO Phase I-III) have been analysed. This blazar-type object was detected by COMPTEL several times along the course of the CGRO-mission. 
Detections and non-detections indicate a time variable MeV-flux. 
While at the highest COMPTEL energies the longterm light curve follows the trend as seen by EGRET (E$>$100~MeV), it does not so at photon energies below 3~MeV, indicating different spectral states of the blazar. 
Combining simultaneously measured COMPTEL and EGRET spectra indicate a bimodal spectral behaviour. During so called 'low' states the combined spectra can be fit with a single power-law representation while during the so called 'high' states a spectral break at energies around $\sim$10~MeV is obvious and therefore requiring models with a spectral turnover at MeV-energies to fit the data. In the 'low' state, only the inclusion of OSSE data requires a low energy spectral break at $\sim$2~MeV. The COMPTEL spectra themselves are also quite different for these two states: a soft spectrum for the 'low' state and a hard spectrum for the 'high' state, with however roughly equal flux levels at energies below 3~MeV. These observational facts would be consistent with the appearance of an additional spectral component during the 'high' or 'flaring' state at energies above 3~MeV mainly. 

Our results fit well into the general relativistic jet scenario for \gray emission of blazars: time variability, maximal energy release at $\gamma$-rays, and a spectral turnover at MeV-energies is observed. However, we found evidence that for \gray flaring phases at least two different spectral components are responsible for the MeV to GeV photon spectrum.

\medskip
{\sf Acknowledgements.} This research was supported by the Deutsche Agentur f\"ur Raumfahrtangelegenheiten (DARA) under the grant 50 QV 90968, by NASA under   contract NASA-26645, and by the Netherlands Organisation for Scientific Research. 

\medskip


\end{document}